\documentclass[%
reprint,
superscriptaddress,
amsmath,amssymb,
aps,
prc,
]{revtex4-2}

\usepackage{graphicx}
\usepackage{dcolumn}
\usepackage{bm}
\usepackage[range-phrase={\,\text{--}\,}, range-units={single}, separate-uncertainty=true]{siunitx}
\usepackage{booktabs}
\usepackage{tablefootnote}
\usepackage{subfigure}

\newcommand{\pg}{\ensuremath{({p},\gamma)}}

\begin{document}

\preprint{APS/123-QED}

\title{A Comparative Analysis of R-Matrix Fitting: ${}^{12}$C\pg{}${}^{13}$N as Test Case}

\author{J.~Skowronski}
 \email{jakub.skowronski@pd.infn.it}
 \affiliation{Dipartimento di Fisica, Università degli Studi di Padova, 35131 Padova, Italy}
 \affiliation{INFN, Sezione di Padova, 35131 Padova, Italy}

\author{D.~Piatti}
 \affiliation{Dipartimento di Fisica, Università degli Studi di Padova, 35131 Padova, Italy}
 \affiliation{INFN, Sezione di Padova, 35131 Padova, Italy}

\author{D.~Rapagnani}
 \affiliation{Dipartimento di Fisica ``E. Pancini'', Universit\`a degli Studi di Napoli ``Federico II'', 80125 Naples, Italy}
 \affiliation{INFN, Sezione di Napoli, 80125 Naples, Italy}

\author{M.~Aliotta}
 \affiliation{SUPA, School of Physics and Astronomy, University of Edinburgh, EH9 3FD Edinburgh, United Kingdom}

\author{C.~Ananna}
 \affiliation{Dipartimento di Fisica ``E. Pancini'', Universit\`a degli Studi di Napoli ``Federico II'', 80125 Naples, Italy}
 \affiliation{INFN, Sezione di Napoli, 80125 Naples, Italy}

\author{L.~Barbieri}
 \affiliation{SUPA, School of Physics and Astronomy, University of Edinburgh, EH9 3FD Edinburgh, United Kingdom}

\author{F.~Barile}
 \affiliation{Dipartimento di Fisica ``M. Merlin'', Università degli Studi di Bari ``A. Moro'', 70125 Bari, Italy}
 \affiliation{INFN, Sezione di Bari, 70125 Bari, Italy}

\author{D.~Bemmerer}
 \affiliation{Helmholtz-Zentrum Dresden-Rossendorf, 01328 Dresden, Germany}

\author{A.~Best}
 \affiliation{Dipartimento di Fisica ``E. Pancini'', Universit\`a degli Studi di Napoli ``Federico II'', 80125 Naples, Italy}
 \affiliation{INFN, Sezione di Napoli, 80125 Naples, Italy}

\author{A.~Boeltzig}
 \affiliation{Helmholtz-Zentrum Dresden-Rossendorf, 01328 Dresden, Germany}
 \affiliation{INFN, Laboratori Nazionali del Gran Sasso, 67100 Assergi, Italy}

\author{C.~Broggini}
 \affiliation{INFN, Sezione di Padova, 35131 Padova, Italy}

\author{C.\,G.~Bruno}
 \affiliation{SUPA, School of Physics and Astronomy, University of Edinburgh, EH9 3FD Edinburgh, United Kingdom}

\author{A.~Caciolli}
 \email{antonio.caciolli@pd.infn.it}
 \affiliation{Dipartimento di Fisica, Università degli Studi di Padova, 35131 Padova, Italy}
 \affiliation{INFN, Sezione di Padova, 35131 Padova, Italy}

\author{M.~Campostrini}
 \affiliation{Laboratori Nazionali di Legnaro, 35020 Legnaro, Italy}

\author{F.~Casaburo}
 \affiliation{INFN, Sezione di Genova, 16146 Genova, Italy}
 \affiliation{Università degli Studi di Genova, 16146 Genova, Italy}

\author{F.~Cavanna}
 \affiliation{INFN, Sezione di Torino, 10125 Torino, Italy}

\author{G.\,F.~Ciani}
 \affiliation{Dipartimento di Fisica ``M. Merlin'', Università degli Studi di Bari ``A. Moro'', 70125 Bari, Italy}
 \affiliation{INFN, Sezione di Bari, 70125 Bari, Italy}

\author{P.~Colombetti}
 \affiliation{Dipartimento di Fisica, Universit\`a degli Studi di Torino, 10125 Torino, Italy}
 \affiliation{INFN, Sezione di Torino, 10125 Torino, Italy}

\author{A.~Compagnucci}
 \affiliation{Gran Sasso Science Institute, 67100 L'Aquila, Italy}
 \affiliation{INFN, Laboratori Nazionali del Gran Sasso, 67100 Assergi, Italy}

\author{P.~Corvisiero}
 \affiliation{Università degli Studi di Genova, 16146 Genova, Italy}
 \affiliation{INFN, Sezione di Genova, 16146 Genova, Italy}

\author{L.~Csedreki}
 \affiliation{HUN-REN Institute for Nuclear Research (ATOMKI), PO Box 51, H-4001 Debrecen, Hungary}

\author{T.~Davinson}
 \affiliation{SUPA, School of Physics and Astronomy, University of Edinburgh, EH9 3FD Edinburgh, United Kingdom}

\author{R.~Depalo}
 \affiliation{Università degli Studi di Milano, 20133 Milano, Italy}
 \affiliation{INFN, Sezione di Milano, 20133 Milano, Italy}

\author{A.~Di Leva}
 \affiliation{Dipartimento di Fisica ``E. Pancini'', Universit\`a degli Studi di Napoli ``Federico II'', 80125 Naples, Italy}
 \affiliation{INFN, Sezione di Napoli, 80125 Naples, Italy}

\author{Z.~Elekes}
 \affiliation{HUN-REN Institute for Nuclear Research (ATOMKI), PO Box 51, H-4001 Debrecen, Hungary}
 \affiliation{Institute of Physics, Faculty of Science and Technology, University of Debrecen, Egyetem tér 1., H-4032 Debrecen, Hungary}

\author{F.~Ferraro}
 \affiliation{INFN, Laboratori Nazionali del Gran Sasso, 67100 Assergi, Italy}

\author{A.~Formicola}
 \affiliation{INFN, Sezione di Roma, 00185 Roma, Italy}

\author{Zs.~Fülöp}
 \affiliation{HUN-REN Institute for Nuclear Research (ATOMKI), PO Box 51, H-4001 Debrecen, Hungary}

\author{G.~Gervino}
 \affiliation{Dipartimento di Fisica, Universit\`a degli Studi di Torino, 10125 Torino, Italy}
 \affiliation{INFN, Sezione di Torino, 10125 Torino, Italy}

\author{R.\,M.~Gesu\`e}
 \affiliation{Gran Sasso Science Institute, 67100 L'Aquila, Italy}
 \affiliation{INFN, Laboratori Nazionali del Gran Sasso, 67100 Assergi, Italy}

\author{A.~Guglielmetti}
 \affiliation{Università degli Studi di Milano, 20133 Milano, Italy}
 \affiliation{INFN, Sezione di Milano, 20133 Milano, Italy}

\author{C.~Gustavino}
 \affiliation{INFN, Sezione di Roma, 00185 Roma, Italy}

\author{Gy.~Gyürky}
 \affiliation{HUN-REN Institute for Nuclear Research (ATOMKI), PO Box 51, H-4001 Debrecen, Hungary}

\author{G.~Imbriani}
 \affiliation{Dipartimento di Fisica ``E. Pancini'', Universit\`a degli Studi di Napoli ``Federico II'', 80125 Naples, Italy}
 \affiliation{INFN, Sezione di Napoli, 80125 Naples, Italy}

\author{M.~Junker}
 \affiliation{INFN, Laboratori Nazionali del Gran Sasso, 67100 Assergi, Italy}

\author{M.~Lugaro}
 \affiliation{Konkoly Observatory, Research Centre for Astronomy and Earth Sciences (CSFK), MTA Centre for Excellence, 1121 Budapest, Hungary}
 \affiliation{ELTE E\"otv\"os Lor\'and University, Institute of Physics and Astronomy, 1117 Budapest, Hungary}

\author{P.~Marigo}
 \affiliation{Dipartimento di Fisica, Università degli Studi di Padova, 35131 Padova, Italy}
 \affiliation{INFN, Sezione di Padova, 35131 Padova, Italy}

\author{J.~Marsh}
 \affiliation{SUPA, School of Physics and Astronomy, University of Edinburgh, EH9 3FD Edinburgh, United Kingdom}

\author{E.~Masha}
 \affiliation{Helmholtz-Zentrum Dresden-Rossendorf, 01328 Dresden, Germany}
 \affiliation{Università degli Studi di Milano, 20133 Milano, Italy}

\author{R.~Menegazzo}
 \affiliation{INFN, Sezione di Padova, 35131 Padova, Italy}

\author{D.~Mercogliano}
 \affiliation{Dipartimento di Fisica ``E. Pancini'', Universit\`a degli Studi di Napoli ``Federico II'', 80125 Naples, Italy}
 \affiliation{INFN, Sezione di Napoli, 80125 Naples, Italy}

\author{V.~Paticchio}
 \affiliation{INFN, Sezione di Bari, 70125 Bari, Italy}

\author{R.~Perrino}
 \altaffiliation[Permanent address: ]{INFN Sezione di Lecce, 73100 Lecce, Italy}
 \affiliation{INFN, Sezione di Bari, 70125 Bari, Italy}

\author{P.~Prati}
 \affiliation{Università degli Studi di Genova, 16146 Genova, Italy}
 \affiliation{INFN, Sezione di Genova, 16146 Genova, Italy}

\author{V.~Rigato}
 \affiliation{Laboratori Nazionali di Legnaro, 35020 Legnaro, Italy}

\author{D.~Robb}
 \affiliation{SUPA, School of Physics and Astronomy, University of Edinburgh, EH9 3FD Edinburgh, United Kingdom}

\author{L.~Schiavulli}
 \affiliation{Dipartimento di Fisica ``M. Merlin'', Università degli Studi di Bari ``A. Moro'', 70125 Bari, Italy}
 \affiliation{INFN, Sezione di Bari, 70125 Bari, Italy}

\author{R.\,S.~Sidhu}
 \affiliation{SUPA, School of Physics and Astronomy, University of Edinburgh, EH9 3FD Edinburgh, United Kingdom}
 \affiliation{School of Mathematics and Physics, University of Surrey, Guildford, GU2 7XH, United Kingdom}

\author{O.~Straniero}
 \affiliation{INAF-Osservatorio Astronomico d'Abruzzo, 64100, Teramo, Italy}
 \affiliation{INFN, Sezione di Roma, 00185 Roma, Italy}

\author{T.~Szücs}
 \affiliation{HUN-REN Institute for Nuclear Research (ATOMKI), PO Box 51, H-4001 Debrecen, Hungary}

\author{S.~Turkat}
 \affiliation{Dipartimento di Fisica, Università degli Studi di Padova, 35131 Padova, Italy}
 \affiliation{INFN, Sezione di Padova, 35131 Padova, Italy}

\author{S.~Zavatarelli}
 \affiliation{INFN, Sezione di Genova, 16146 Genova, Italy}
 \affiliation{Università degli Studi di Genova, 16146 Genova, Italy}

\collaboration{LUNA Collaboration}
\noaffiliation


\date{\today}

\begin{abstract}
In nuclear astrophysics, the accurate determination of nuclear reaction cross sections at astrophysical energies is critical for understanding stellar evolution and nucleosynthesis. This study focuses on the $^{12}$C($p, \gamma$)$^{13}$N reaction, which takes part in the CNO cycle and is significant for determining the $^{12}$C/$^{13}$C ratio in stellar interiors. Data from various studies, including recent LUNA measurements, reveal high discrepancies in cross section values, underscoring the need for robust fitting approaches. Utilizing the R-matrix theory, we compare different frequentist and Bayesian methodologies for estimating reaction cross sections and their uncertainties. The analysis evaluates the strengths and weaknesses of different statistical techniques, highlighting the importance of systematic uncertainty treatment and the estimate of covariance matrix estimation to enhance the reliability and reproducibility of uncertainty estimates in nuclear astrophysics.
\end{abstract}

\maketitle


\section{\label{sec:intro}Introduction}

In the field of nuclear astrophysics, the precise determination of nuclear reaction cross sections at astrophysical energies is of paramount significance for understanding stellar objects, the origin of elements and the evolution of our Universe. In the last three decades, the LUNA collaboration has been attempting to measure these quantities as close as possible to the energies relevant for the astrophysical scenarios where they occur. This was possible due to the special environment provided by the Laboratori Nazionali del Gran Sasso (LNGS), where the cosmic background is drastically reduced~\cite{broggini2018}. However, in many cases this approach is still not enough to cover all the existing scenarios. Hence, the knowledge of these cross sections often relies on the extrapolation of data obtained in laboratory experiments to the low-energy region relevant to the astrophysical environments by the mean of R-matrix theory \cite{descouvemont2010}. It offers the possibility of exploiting in a meaningful way all the experimental information about the reaction, such as measured cross sections, asymptotic normalization coefficients (ANC), angular distributions and resonance proprieties. A significant challenge in this endeavor lies in quantifying the associated uncertainties to these extrapolations, as there are still no well formalized and accepted methodologies for this. Traditionally, the frequentist approaches have been employed~\cite{deboer2017}. However, in the recent years, the Bayesian method started to emerge~\cite{odell2022}, both due to the fact that it started to be computationally accessible and because it provides a natural way of sampling the entire parameter space. Nevertheless, the two approaches were never systematically studied nor compared to each other within the context of R-matrix fitting, making their strengths, weaknesses and applicability unclear. The R-matrix has been frequently used to estimate the reaction rates of several different nuclear reactions, always struggling with the uncertainty estimation. As an example, the ${}^{13}$C\pg{}${}^{14}$N reaction was carefully studied~\cite{chakraborty2015}, however, the uncertainty of the fit was simply taken as the systematic uncertainty of the main dataset. The most recent study of the ${}^{14}$N\pg{}${}^{15}$O reaction~\cite{frentz2022} discussed about the large discrepancies between the datasets and provided fruitful insights on the different choices that have been made, but gave no precise information on how the uncertainty was estimated. Additionally, the covariance matrices between the different parameters, which show how the different parameters interact with each other and greatly affects the uncertainty estimation, were never properly studied nor estimated in the known literature. Hence, this paper aims to present a comparative analysis of different methodologies and their implications to obtain reliable and reproducible estimates of uncertainty in the R-matrix fitting for nuclear astrophysics.

As case study we choose the ${}^{12}$C\pg{}${}^{13}$N reaction. This reaction takes part in the CNO cycle inside stellar interiors and governs, alongside the ${}^{13}$C\pg{}${}^{14}$N reaction, the $^{12}$C/$^{13}$C ratio in the hydrogen burning regions. This, in turn, influences the $^{12}$C/$^{13}$C ratio in the stellar atmosphere, making it a useful parameter to asses the stage of stellar nucleosynthesis. In fact, this isotopic ratio can be used to test mixing in the Red Giant Branch (RGB) and Asymptotic Giant Branch (AGB) stars, whose occurance and origin is still under debate~\cite{karakas2014,palmerini2021}. Additionally, the ${}^{12}$C\pg{}${}^{13}$N reaction indirectly induces neutrino production in stars through the $\beta ^{+}$ decay of the unstable ${}^{13}$N, making it significant for the calculation of the the neutrino flux from our Sun~\cite{bahcall2006}. This reaction was recently measured by several research groups~\cite{skowronski2023_prc,skowronski2023_prl,gyurky2023,kettner2023, csedreki2023}, showing an important discrepancy with respect to the previous studies~\cite{vogl1963,rolfs1974,baily1950,lamb1957,burtebaev2008} and pointing towards a cross section systematically lower than previously thought. Due to its significance and discrepancy, togheter with the fact that the ${}^{12}$C\pg{}${}^{13}$N reaction was already used in the past as a R-matrix code benchmark \cite{azuma2010}, it is a good test case for searching a robust R-matrix fit approach and clear procedures for the uncertainty estimation.

\section{\label{sec:methods}Analysis}

\subsection{Nuclear Data Input}

\begin{figure}
    \centering
    \includegraphics[width=0.95\linewidth]{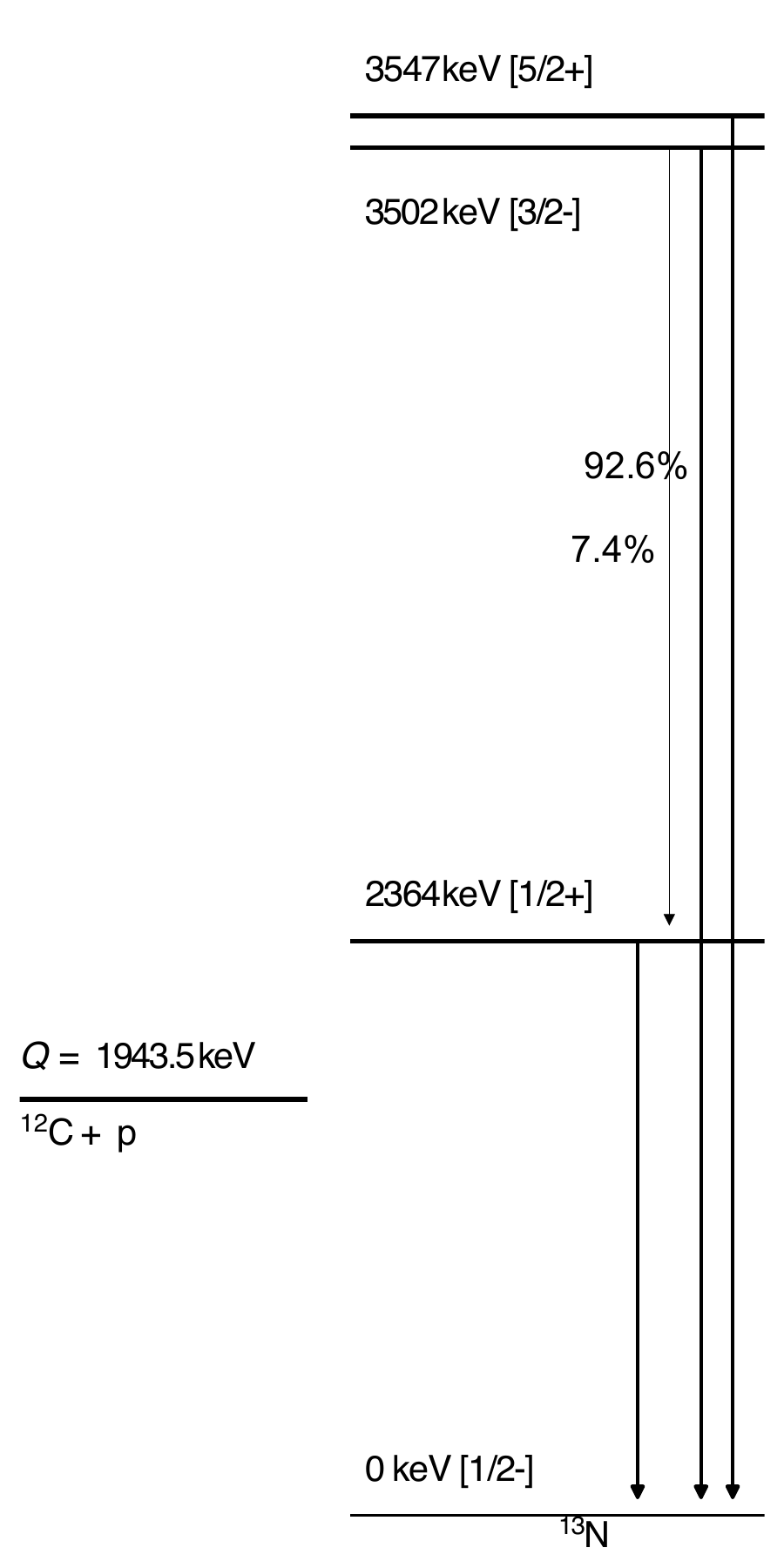}
    \caption{The level scheme of the $^{13}$N nucleus.}
    \label{fig:levels}
\end{figure}

To construct the R-matrix framework for the ${}^{12}$C\pg{}${}^{13}$N reaction, first we collected all the information about the $^{13}$N nucleus and the connected reaction channels. In the energy window of interest of the $(p,\gamma)$ channel, i.e.\ \SIrange[]{0}{2}{\mega\electronvolt}, the $^{13}$N nucleus exhibits three excited states, namely \SI{2.368}{\mega\electronvolt},  \SI{3.503}{\mega\electronvolt} and \SI{3.545}{\mega\electronvolt}, corresponding to three resonances in the radiative capture process. These states decay mainly to the ground state through E1, M1 and M2 transitions, respectively, with an additional \SI{7.4}{\percent} E1 transition of the \SI{3.503}{\mega\electronvolt} to the \SI{2.368}{\mega\electronvolt} state. The level scheme and the $Q$-value of the reaction are shown in Fig.~\ref{fig:levels}.
To include the effects of the low energy tails of higher energy resonances, it is common in the R-Matrix analysis to include for each spin channel a higher energy state, e.g.\ at \SI{15}{\mega\electronvolt}. For the direct capture component, instead, the external capture integrals are calculated \cite{descouvemont2010}. Hence, both the background poles and the E1 and E2 external contributions were included in the calculation, which were necessary to model the cross section around the region of the second resonance highly dependent on the interference terms.
In the calculations, only the $(p,p)$ and the $(p,\gamma)$ channels were considered, since no other channel is open at the energies of interest. The channel radius was set to \SI{3.6}{fm}, with a separation energy of \SI{1.9435}{\mega\electronvolt} \cite{NSR2012WA38}. The initial partial widths for each channel were taken from the most recent R-matrix study~\cite{kettner2023}. For the \SI{3.545}{\mega\electronvolt} resonance, no radiative width was considered since there is no experimental evidence in the literature of its contribution to the radiative capture cross section \cite{rolfs1974, kettner2023}. However, the particle width was included for a correct evaluation of the elastic channel, which shows evidence of the \SI{3.503}{\mega\electronvolt} and \SI{3.545}{\mega\electronvolt} doublet \cite{meyer1976}. The values for each partial width and the resonance positions are shown in Table~\ref{tab:initial}. The ground state asymptotic normalization constant (ANC) was taken from~\citet{artemov2022}.

\begin{table*}[htb]
\centering
\caption{R-matrix parameters for the ${}^{12}$C\pg{}${}^{13}$N reaction taken from the latest study~\cite{kettner2023}, except for the ANC for the ground state from~\cite{artemov2022}. The $\Gamma _{\gamma}$ sign of \SI{3.503}{\mega\electronvolt} resonance was changed since it was not possible to reproduce the~\citet{kettner2023} fit with a positive one.}
\begin{tabular}{c|c|c|c|c}
\toprule
\toprule
$E_{p}$ (MeV) & $E_{x}$ (MeV) & $J^{\pi}$ & $\Gamma _{p}$ (keV) or ANC (fm$^{-1/2}$) & $\Gamma _{\gamma}$ (eV) \\ 
\midrule
\midrule
& \num{0} & 1/2$^{-}$  & \num{1.62 \pm 0.5}  &  \\
\num{0.4603 \pm 0.0005}  & \num{2.3682 \pm 0.0005} & 1/2$^{+}$   & \num{34.0 \pm 0.2}  & \num{-0.48 \pm 0.3} \\
\num{1.6888 \pm 0.0005} & \num{3.5032 \pm 0.0005} & 3/2$^{-}$   & \num{55.2 \pm 0.3}  & \num{-0.49 \pm 0.3} / (\num{7.2 \pm 1.1})$\times 10^{-4}$ \footnotemark[2] \\
\num{1.7355 \pm 0.0005} & \num{3.5453 \pm 0.0005} & 5/2$^{+}$   & \num{49.0 \pm 0.5}  & \\
& \num{20} \footnotemark[1] & 1/2$^{+}$   & $5 \times 10^{3}$   & (\num{5.4 \pm 0.5}) $\times 10^{3}$ \\
& \num{20} \footnotemark[1] & 3/2$^{-}$   & $5 \times 10^{3}$   & \num{-120 \pm 20} / \num{1.1 \pm 0.2}$\times 10^{3}$ \footnotemark[2] \\
\bottomrule
\bottomrule
\end{tabular}
\label{tab:initial}
\footnotetext[1]{Background pole}
\footnotetext[2]{M1/E2 transitions}
\end{table*}

\subsection{State of the Art}

\begin{figure*}
    \centering
    \includegraphics[width=\linewidth]{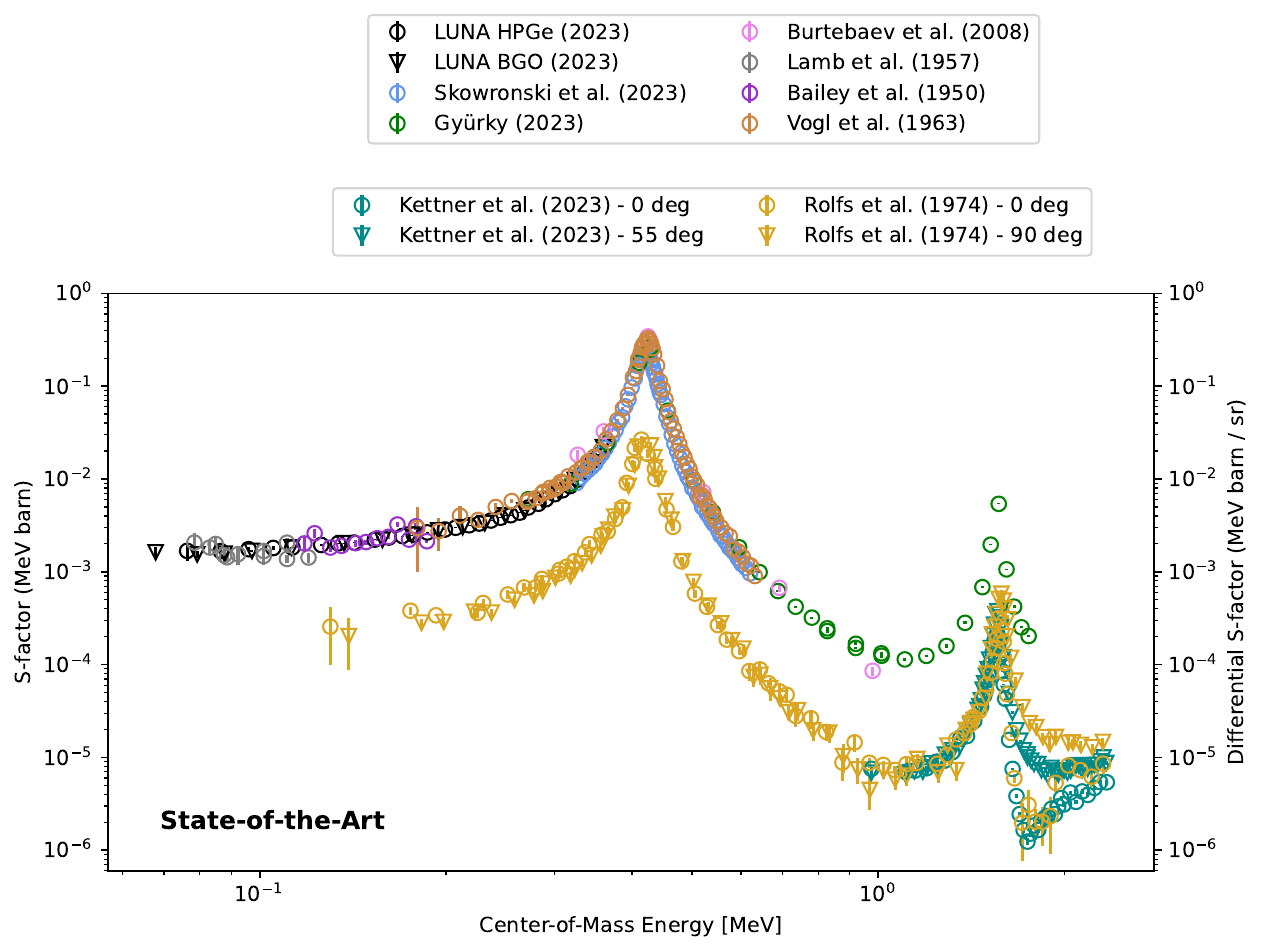}
    \caption{Literature data for the S-factor of the ${}^{12}$C\pg{}${}^{13}$N reaction. The \cite{rolfs1974} and \cite{kettner2023} datasets are presented in form of differential cross sections, as reported in respective publications. "LUNA" refers to \cite{skowronski2023_prl}.}
    \label{fig:lit}
\end{figure*}

All the literature data for the $(p,\gamma)$ channel~\cite{lamb1957,baily1950,vogl1963,rolfs1974,burtebaev2008,gyurky2023,skowronski2023_prc,skowronski2023_prl,kettner2023} are shown in Fig.~\ref{fig:lit}. \citet{baily1950} and \citet{lamb1957} are the oldest studies, which counted the emitted positrons from the $^{13}$N decay. They spanned the energies from \SI{88}{\kilo\electronvolt} to \SI{200}{\kilo\electronvolt}. Large statistical uncertainties were reported with no clear systematic uncertainty treatment. Hence, they do not strongly constrain the cross section, although they were the first approaches for obtaining the reaction cross section near the astrophysically relevant region. \citet{vogl1963} and \citet{rolfs1974} are extensive studies representing the main references until a few years ago. The former concentrated on the study of the \SI{2.368}{\mega\electronvolt} resonance by detecting the $\gamma$-rays from the direct capture process. The latter spanned the energy window of the \SI{2.368}{\mega\electronvolt} and the \SI{3.503}{\mega\electronvolt} resonances and provided the data in form of the differential cross section at two different angles. No systematic uncertainty is reported in either study. An energy difference of \SI{5}{\kilo\electronvolt} for the position of \SI{2.368}{\mega\electronvolt} resonance is found between the two studies. The datasets, however, agree well with one another in terms of absolute cross section. 
More recently, the \citet{burtebaev2008} study used the yield on a thick carbon disk to extract seven integrated cross section data points in the region around the first resonance. It agrees well with the previous results and is the first study that reports a systematic uncertainty of the data. \citet{gyurky2023} studied the reaction with the activation technique i.e.\ by counting the \SI{511}{\kilo\electronvolt} $\gamma$-rays from the $^{13}$N decay, and spanned a large energy range that includes both resonances in one complete dataset. Their cross section agree with the \citet{vogl1963} and \citet{rolfs1974} results. Nevertheless, the study observed a shift in the position of \SI{3.503}{\mega\electronvolt} resonance of about \SI{25}{\kilo\electronvolt} with respect to \citet{rolfs1974}, which was conducted in two different laboratories, affected by two different energy shifts. In the \citet{skowronski2023_prl} study (referred as LUNA), the reaction was measured both with the prompt-$\gamma$ detection and the activation counting, by using two different detection setup: a HPGe detector in close geometry, and an almost 4$\pi$ BGO detector~\cite{skowronski2023_JPhG}. The two approaches gave consistent results. The study reported at lowest energies up to date, entering in the Gamow Window for both RGB and AGB stars, and show a significant deviation of about \SI{30}{\percent} from the \citet{vogl1963} and \citet{rolfs1974} results. The \citet{skowronski2023_prc} study was the extension of the previous measurement at LUNA \cite{skowronski2023_prl}, with an independent experimental setup. It extended the LUNA results at higher energies with \SI{60}{\kilo\electronvolt} of overlap to check the consistency between the two cross sections. Prompt-$\gamma$ rays were detected with HPGe detectors at several different angles. The study also evidences the presence of a normalization issue with respect to the \citet{vogl1963}, \citet{rolfs1974} and \citet{burtebaev2008} datasets in the region of the \SI{2.368}{\mega\electronvolt} resonance. 
Finally, \citet{kettner2023} study explored the region of the \SI{3.503}{\mega\electronvolt} resonance and also found consistently lower cross section than \cite{rolfs1974}. The results are presented in form of differential cross section at two different angles. Additionally, an energy shift of about \SI{24}{\kilo\electronvolt} in the position of the \SI{3.503}{\mega\electronvolt} resonance was observed with respect to \citet{rolfs1974}, in agreement with \citet{gyurky2023}. It is noted that the absolute scale of both the \citet{vogl1963} and \citet{rolfs1974} studies is wrong due to the incorrect stopping power values that were used and it suggest re-normalizing the data down by \SI{30}{\percent} \cite{kettner2023}. Nevertheless, \citet{gyurky2023} and \citet{burtebaev2008} still show higher cross sections than \cite{skowronski2023_prc,kettner2023}, even when the correct stopping power values are used. For what regards the $(p,p)$ channel, the \citet{meyer1976} data were used since they cover the entire energetic region of interest at 3 different angles.

All the data, apart from LUNA (2023) \cite{skowronski2023_prl}, \citet{skowronski2023_prc} and \citet{kettner2023} which were taken from supplemental material of respective publications, were retrieved from EXFOR database. As suggested by \citet{kettner2023}, the \citet{vogl1963} and \citet{rolfs1974} data were scaled down by \SI{30}{\percent}.

Four fitting methodologies were applied to the collected data, one purely frequentist, one hybrid and two Bayesian ones. Each method is discussed in the following sections. In all calculations, the AZURE2 code was used to conduct R-matrix calculations while the minimization tasks relies on programs developed for the present work and available \cite{pyazr,github}. 


\subsection{Frequentist Approach}

In the first two procedures relying on the frequentist methods the best-fit parameters were found by minimizing the following cost function~\cite{dagostini1994}:

\begin{align}
    \chi ^{2} &= \sum _{i,j} \left[ \left(\frac{n_{j}y_{\textup{obs,i}} - y_{\textup{theo}}}{n_{j}\sigma_{\textup{stat},i}} \right)^{2} + \left(\frac{1 - n_{j}}{\sigma_{\textup{sist},j}}\right)^{2} \right],
\end{align}

where $y_{\textup{obs,i}}$ are the observed cross sections, $y_{\textup{theo}}$ the theoretical predictions, $\sigma_{\textup{stat},i}$ are the statistical errors, $\sigma_{\textup{sist},j}$ the systematic errors associated to each dataset and $n_{j}$ their normalization factors. The $i$-index runs over all the points in a given $j$-index dataset. In the studies that report no systematic uncertainties, as in \cite{vogl1963,rolfs1974,baily1950, lamb1957}, the second term was omitted. However, the normalization parameters were still introduced and left free to vary during the fit. The normalizations are needed for the simultaneous fit of different datasets whose absolute values might disagree and thus spuriously increasing the $\chi^{2}$ value. The best-fit parameter errors for the frequentist approach indirectly depends on the assumption that the reduced chi-squared, $\chi^{2}_{\textup{dof}}$, is close to $1$. This is due to the following equation of the $\sigma_{p}$ of the parameter set $p$:

\begin{align}
    |\chi^{2} \left(p_{\textup{best}} + \sigma_{p}\right) - \chi^{2}\left(p_{\textup{best}}\right)| &< \Delta \chi ^{2} (p). \label{eq:chi2}
\end{align}
where $\Delta \chi ^{2}$ is a number obtained from the calculation of the $1\sigma$ interval given a $\chi ^{2} (p)$ distribution with the number of parameters~\cite{avni1976,cline1970}. In case of the present analysis, with 28 free parameters, a value of 30.1 is obtained.

Thus, if the $\chi^{2}_{\textup{dof}}\left(p_{\textup{best}}\right)$ is bigger than $1$, which means the data are more than $1\sigma$ away from the best-fit, a small variation of the parameters is enough to increase the $\chi ^{2}$ by a given $\Delta \chi ^{2} (p)$. The opposite happens when the reduced chi-squared is lower than $1$. Usually it is noted that in the former case the uncertainties of the data are underestimated, and in the latter they are overestimated~\cite{andrae2010}. 


\begin{figure*}[htp]
  \centering
  \subfigure[Reduced chi-squared values obtained by starting the minimization from different starting points.]{\includegraphics[width=0.48\linewidth]{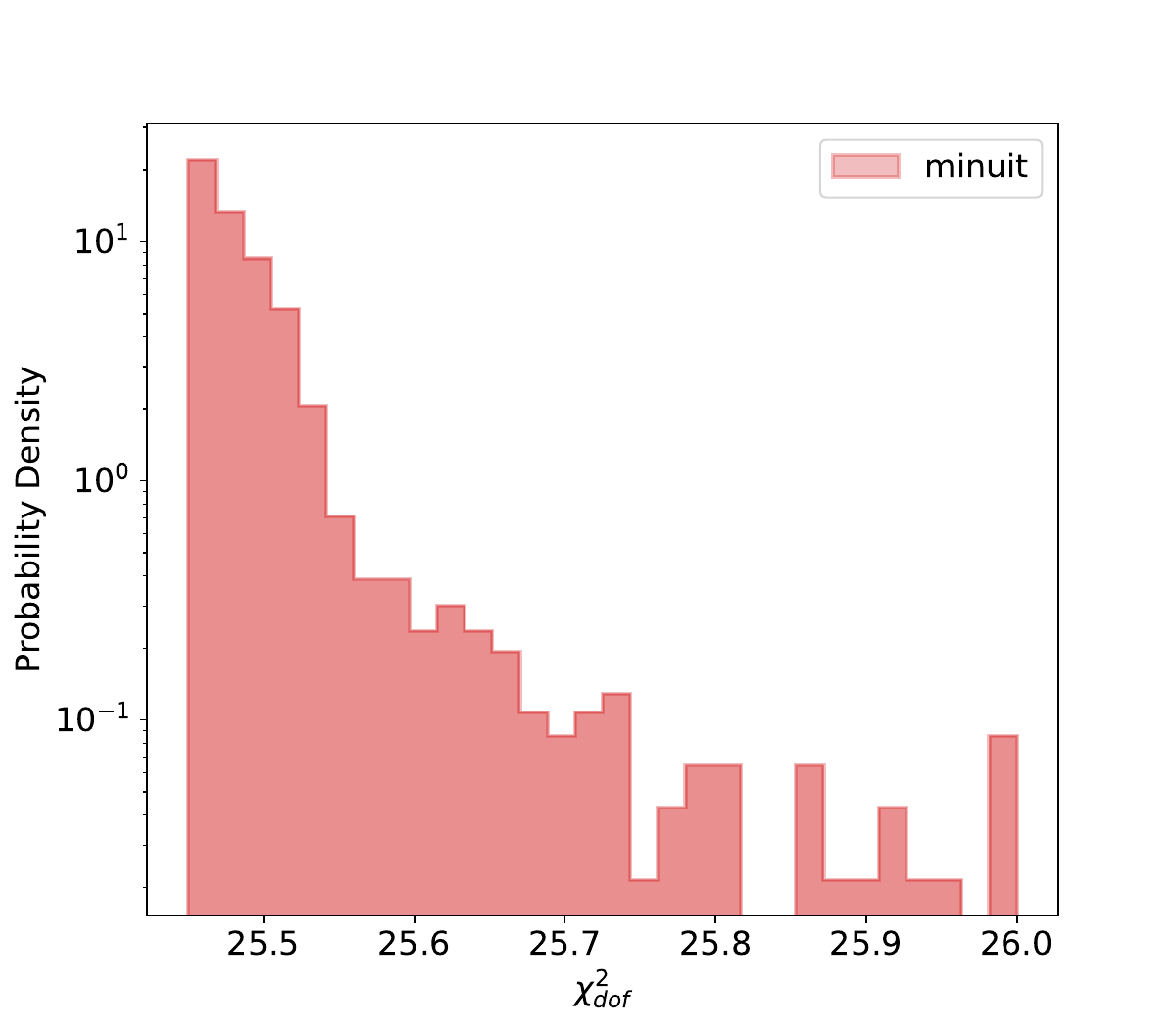}\label{fig:minima}}\quad
  \subfigure[The observed distribution of the LUNA \cite{skowronski2023_prl} dataset best-fit normalization compared with the reported systematic uncertainty.]{\includegraphics[width=0.48\linewidth]{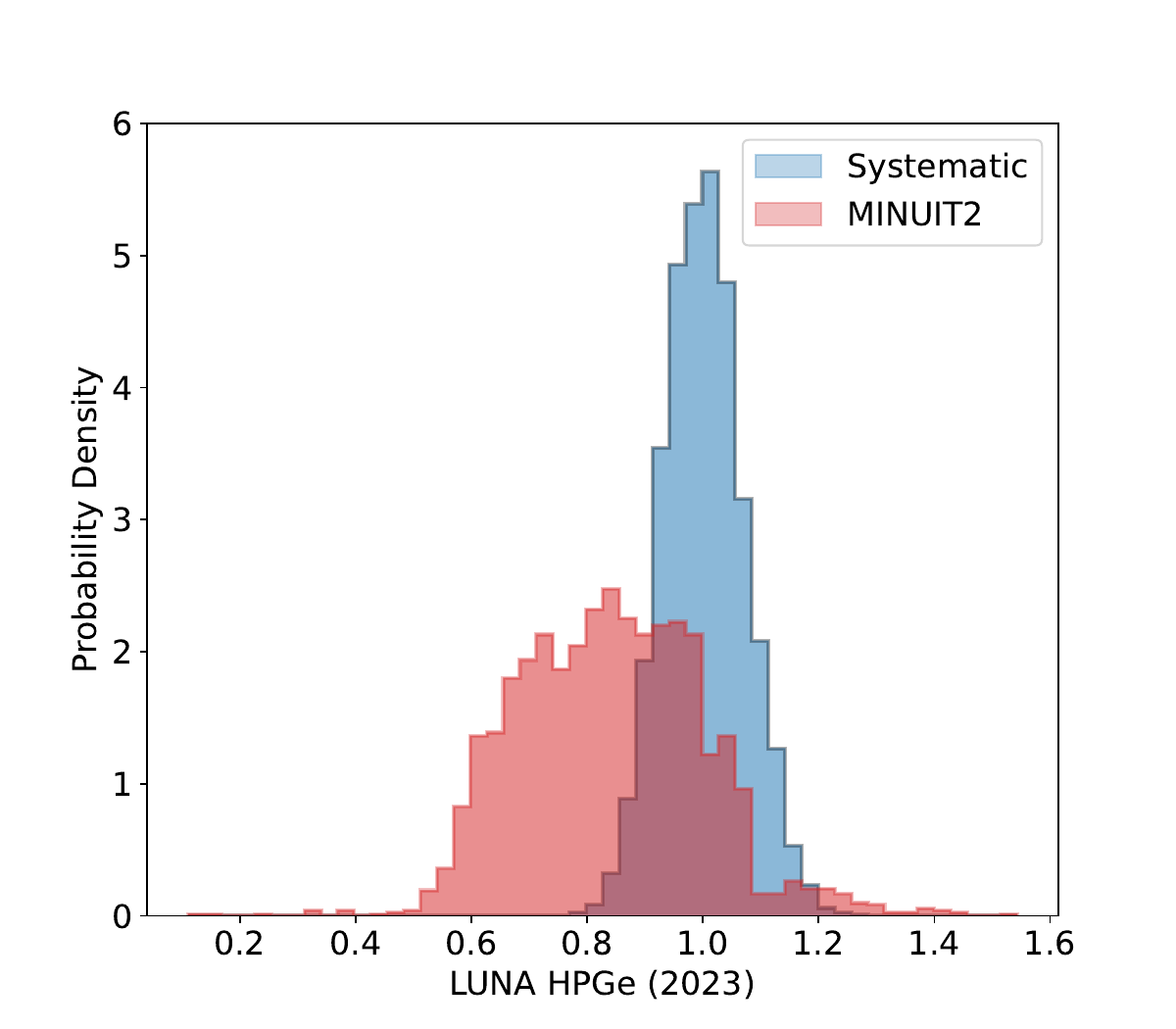}\label{fig:norm-luna}}
  \label{fig:minimizer}
  \caption{The $\chi ^{2} _{\textup{dof}}$ values (panel a) and the LUNA \cite{skowronski2023_prl} normalization parameters (panel b) obtained with MINUIT2 minimizer by changing the initial parameters. The minimizer finds much wider range of values with respect to the reported systematic uncertainty even though comparable chi-squared values are found.}
\end{figure*} 

\begin{figure*}[htp]
  \centering
  \begin{minipage}[b]{0.49\textwidth}
    \subfigure[Residuals for the integrated ($p,\gamma$) cross section.]{\includegraphics[width=1\linewidth]{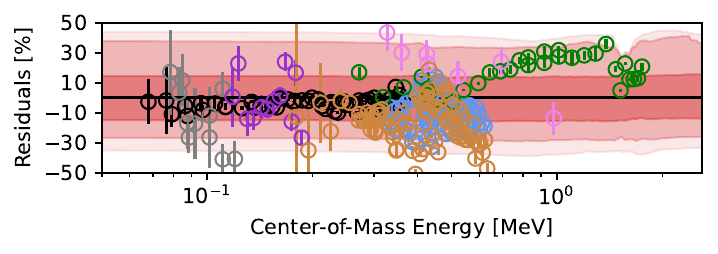}\label{fig:init-1}}\quad
    \subfigure[Residuals for the differential ($p,\gamma$) cross section at 0 deg.]{\includegraphics[width=1\linewidth]{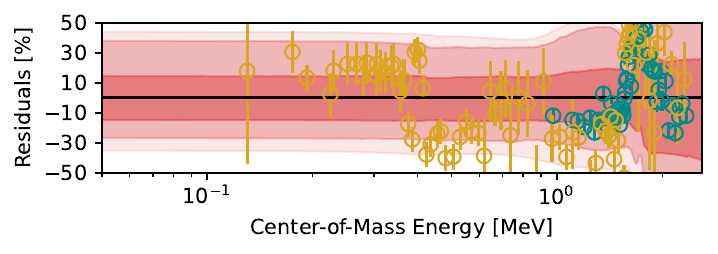}\label{fig:init-2}}
  \end{minipage}
  \subfigure[The correlation matrix among the parameters]{\includegraphics[width=0.48\linewidth]{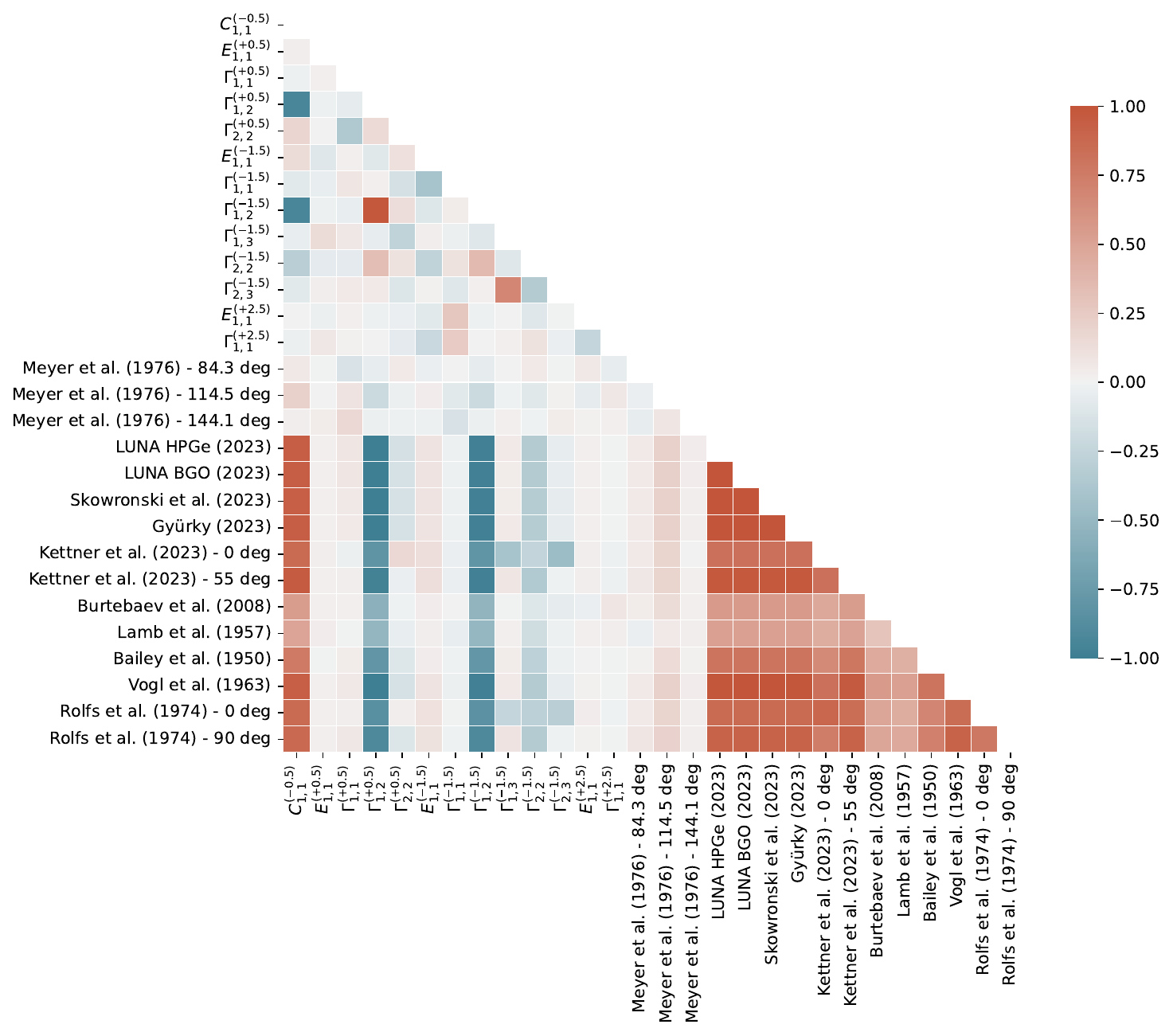}\label{fig:init-3}}
  \label{fig:init}
  \caption{The residuals obtained with the classic frequentist approach (panel a and b). The filled regions represents, respectively, the 68\%, 95\% and 99\% confidence percentiles. The error band is uniform through the spanned energies. The correlation matrix (panel c) shows high values for all the normalization parameters. The data without reported systematic uncertainty are normalized to the best fit values. The data legend is reported in Fig.~\ref{fig:lit}.}
\end{figure*}

The frequentist minimization was initially performed with the use of MINUIT2 \cite{hatlo2005}. However, it was observed that the minimizer tended to give quite different best-fit parameters with similar $\chi ^{2}$ values, depending on the starting position of the parameters. To study this, the initial parameters were sampled from a \SI{50}{\percent} region around the values in Tab. \ref{tab:initial} and the minimization was performed 1000 times. The result can be seen in Fig. \ref{fig:minima} and Fig. \ref{fig:norm-luna} where the $\chi ^{2} _{\textup{dof}}$ values are shown and the distribution of one of the normalization parameters is compared with the nominal systematic uncertainty. This behaviour can be explained by the presence of comparable local minima caused by the second term in Equation \ref{eq:chi2}. Since this term is proportional to the number of datasets, whereas the first term is proportional to the number of data points, the former dominates the $\chi ^{2}$, while the latter readjusts the absolute scale of the cross sections without significantly affecting the value of the $\chi ^{2}$, thus creating many different local minima where the minimizing algorithms hangs. To fix this issue, the \textit{lmfit} package~\cite{lmfit} was used instead of MINUIT2. More precisely, the default Levenberg-Marquardt algorithm \cite{more1977} was selected. The same check was performed as in case of the MINUIT2 minimizer, and in most cases the same best-fit values were found. 
Then, a problem arose with the covariance matrix, which estimates the error of the parameters. Its estimate from the minimizer was showing very small values due to the fact that the $\chi^{2}_{\textup{dof}}$ value was close to $25$, higher than the expected $1$. To overcome this difficulty, the uncertainties of all the cross sections were inflated by a factor of $\sqrt{\chi^{2}_{\textup{dof}}}$, as recommended in \cite{adelberger2011}, including the elastic scattering data. The scaling factor amounted to $5$ and was used only for the current analysis. An alternative could have been scaling the uncertainties with the weight given by the partial contribution of each data set in the total $\chi ^{2}$, but this arguably could introduce a bias in the result. Thus only a unique scaling factor was indiscriminately applied. The uncertainty was then calculated by sampling a multivariate normal distribution constructed from the covariance matrix of the parameters. The resulting residuals are shown in Fig.~\ref{fig:init-1}~-~\ref{fig:init-3}, with the correlation matrix between the parameters. The latter is clearly governed by the normalization parameters, showing a correlation close to unity, which makes the uncertainty estimation unreliable \cite{minuit}. The figure also evidences the issue with the presence of several local minima: since the absolute scale of the data governs the final result of the fit and the normalization parameters are not strongly constrained, as mentioned before, the scale of the best-fit values is not well constrained. Additionally, the uncertainty obtained by scaling the uncertainties of the data shows no structure and is rather constant throughout all the energy range.

\begin{figure*}[htp]
  \centering
  \begin{minipage}[b]{0.49\textwidth}
    \subfigure[Residuals for the integrated ($p,\gamma$) cross section.]{\includegraphics[width=1\linewidth]{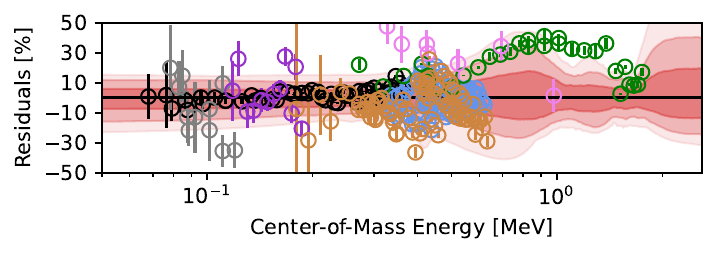}\label{fig:norm-1}}\quad
    \subfigure[Residuals for the differential ($p,\gamma$) cross section at 0 deg.]{\includegraphics[width=1\linewidth]{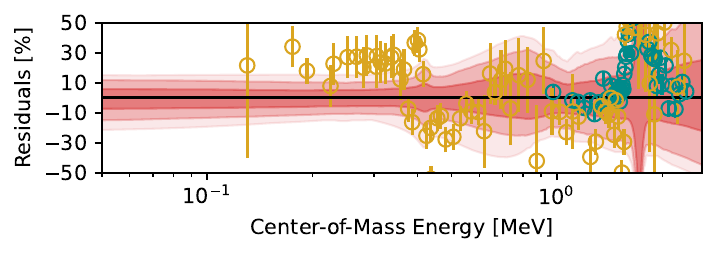}\label{fig:norm-2}}
  \end{minipage}
  \subfigure[The correlation matrix among the parameters]{\includegraphics[width=0.48\linewidth]{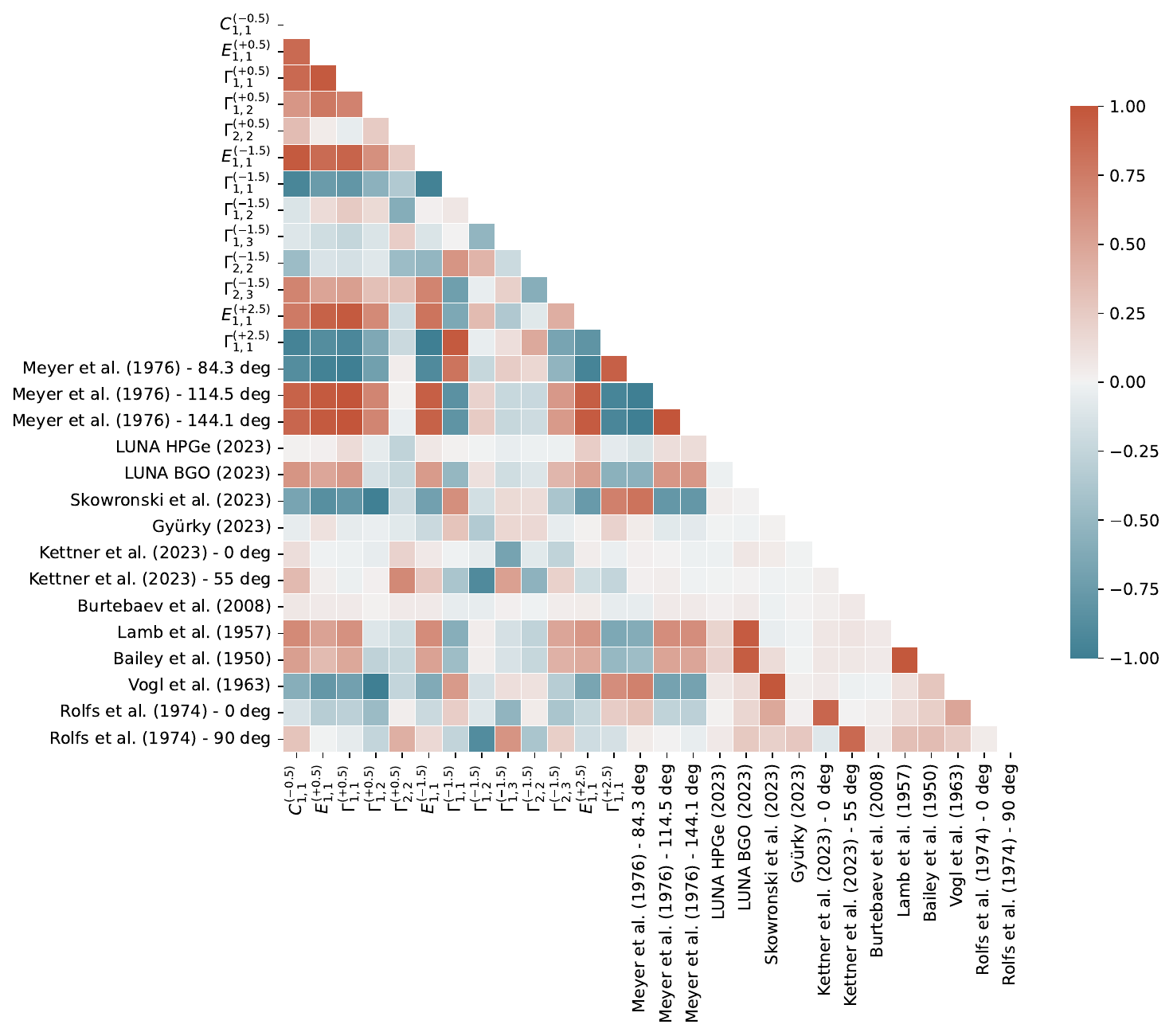}\label{fig:norm-3}}
  \label{fig:norm}
  \caption{The residuals obtained by initially sampling the normalization parameters and then fixing them during the fitting (panel a and b). The filled regions represents, respectively, the 68\%, 95\% and 99\% confidence percentiles. The uncertainty band shows more structure in function of the discrepancies. The correlation matrix (panel c) shows milder correlation, underlining the correlation between the different resonance parameters. The data without reported systematic uncertainty are normalized to the best fit values. The data legend is reported in Fig.~\ref{fig:lit}.}
\end{figure*}

\begin{figure*}[htp]
  \centering
  \begin{minipage}[b]{0.49\textwidth}
    \subfigure[Residuals for the integrated ($p,\gamma$) cross section.]{\includegraphics[width=1\linewidth]{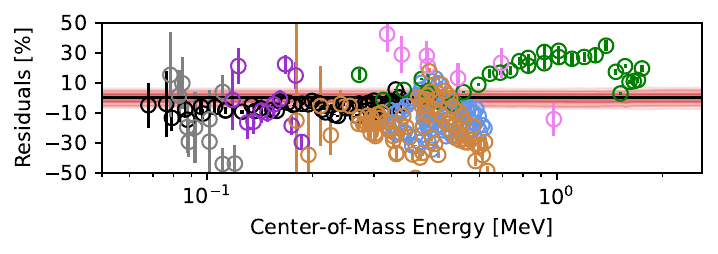}\label{fig:emcee-1}}\quad
    \subfigure[Residuals for the differential ($p,\gamma$) cross section at 0 deg.]{\includegraphics[width=1\linewidth]{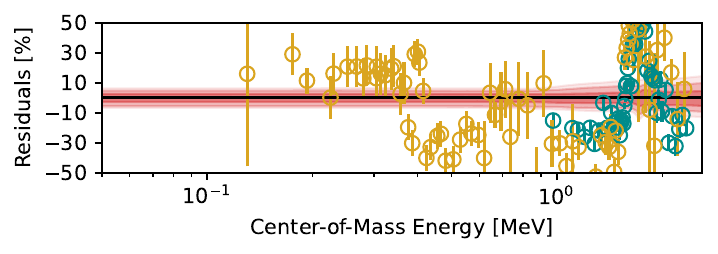}\label{fig:emcee-2}}
  \end{minipage}
  \subfigure[The correlation matrix among the parameters]{\includegraphics[width=0.48\linewidth]{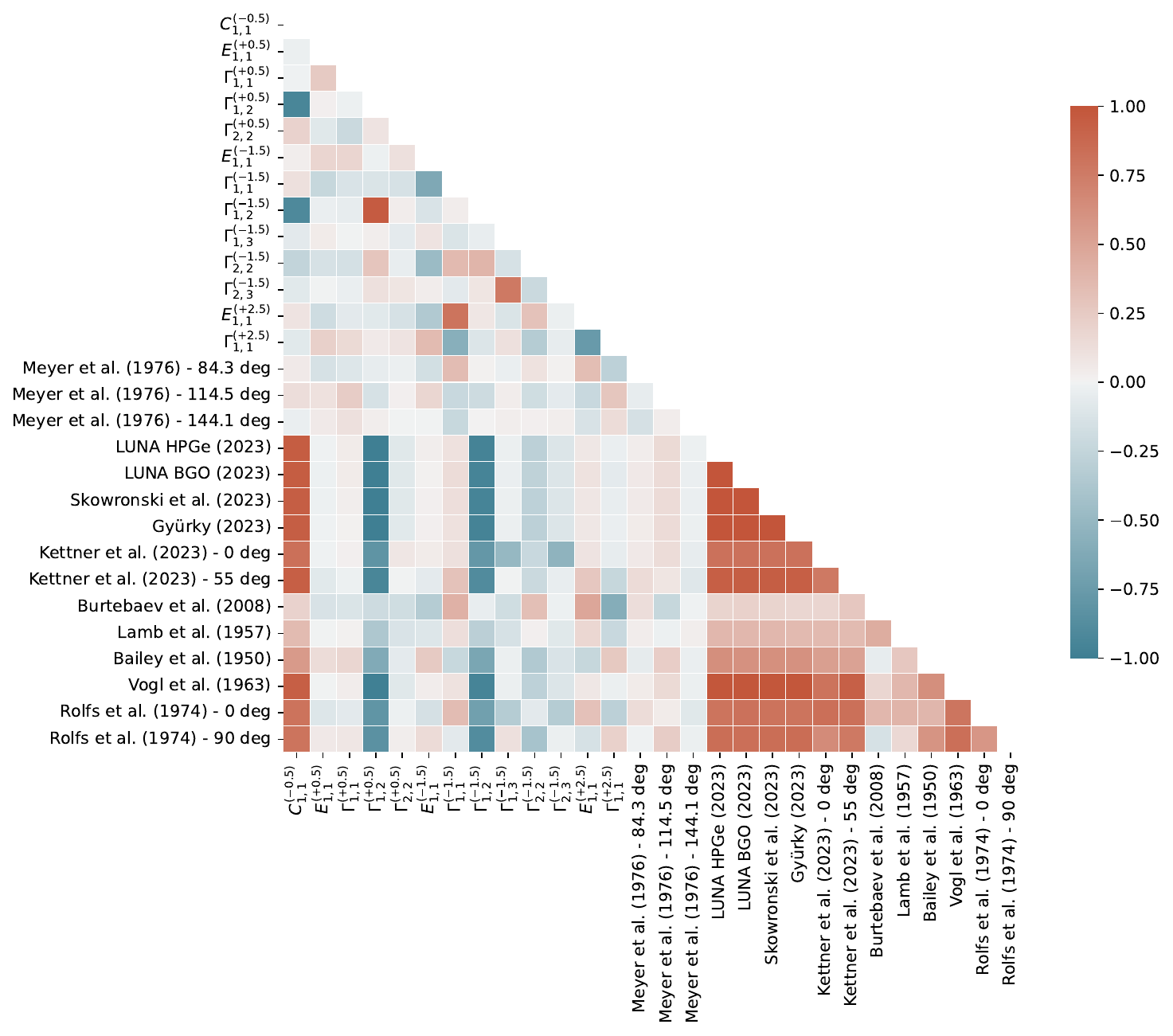}\label{fig:emcee-3}}
  \label{fig:emcee}
  \caption{The residuals obtained by following the Bayesian approach through the Markov Chain Monte Carlo (panel a and b). The filled regions represents, respectively, the 68\%, 95\% and 99\% confidence percentiles. The uncertainty band is smaller with respect to both the frequentist approaches and is constant in the entire energy range. The correlation matrix (panel c) shows large correlation, similar to the situation in Fig.~\ref{fig:init}. The data without reported systematic uncertainty are normalized to the best fit values. The data legend is reported in Fig.~\ref{fig:lit}.}
\end{figure*}

\begin{figure*}[htp]
  \centering
  \begin{minipage}[b]{0.49\textwidth}
    \subfigure[Residuals for the integrated ($p,\gamma$) cross section.]{\includegraphics[width=1\linewidth]{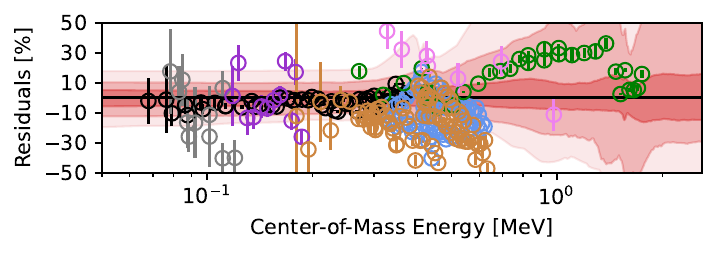}\label{fig:ptemcee-1}}\quad
    \subfigure[Residuals for the differential ($p,\gamma$) cross section at 0 deg.]{\includegraphics[width=1\linewidth]{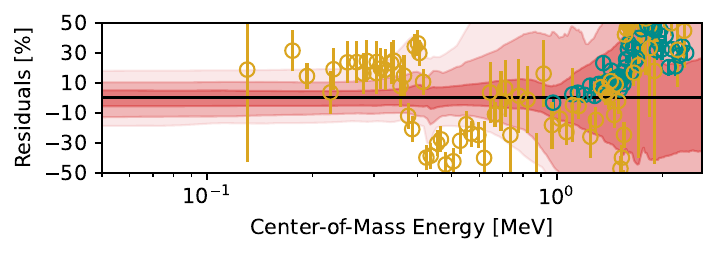}\label{fig:ptemcee-2}}
  \end{minipage}
  \subfigure[The correlation matrix among the parameters]{\includegraphics[width=0.48\linewidth]{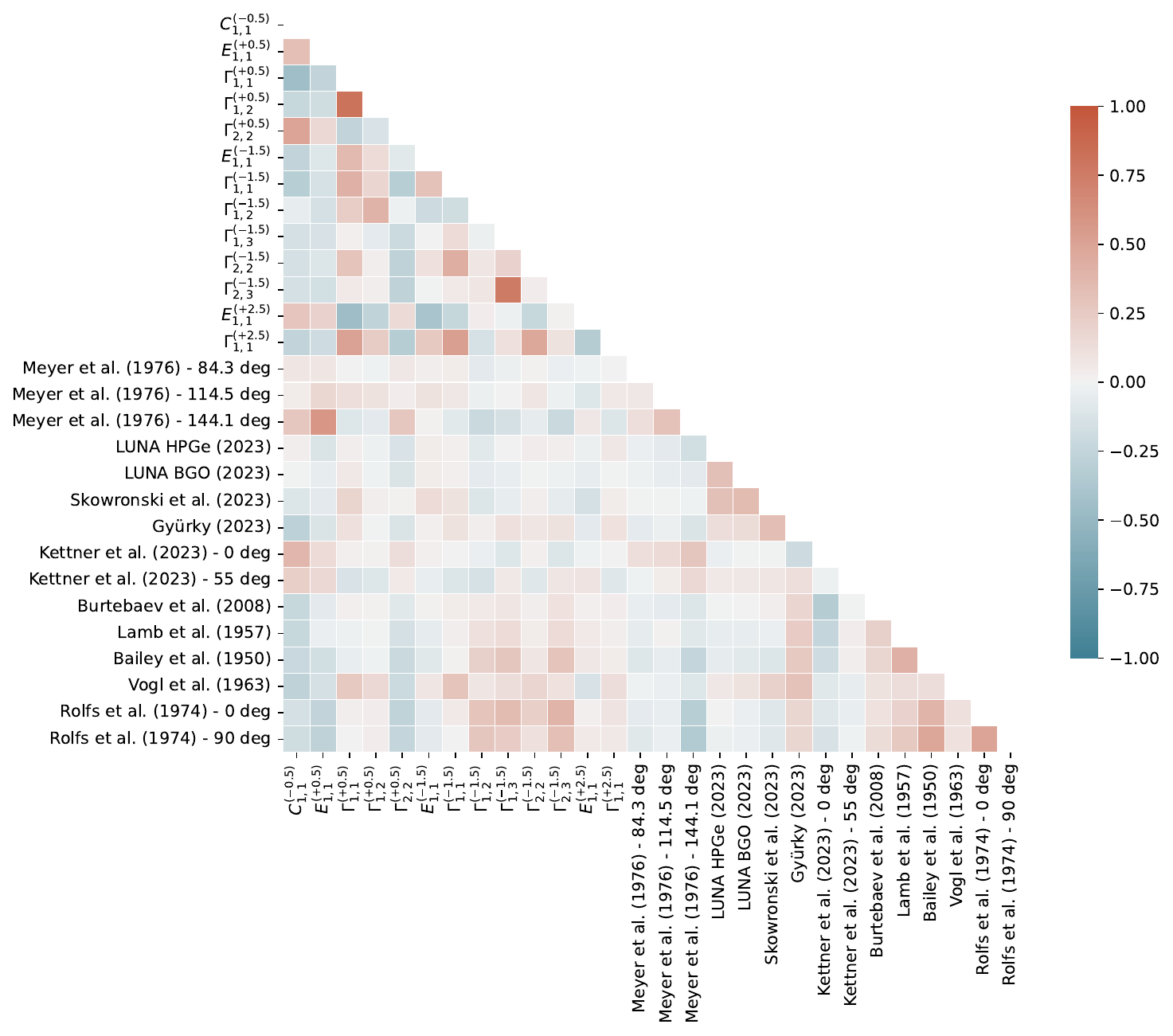}\label{fig:ptemcee-3}}
  \label{fig:ptemcee}
  \caption{The residuals obtained by following the Bayesian approach through the Tempered Markov Chain Monte Carlo (panel a and b). The filled regions represents, respectively, the 68\%, 95\% and 99\% confidence percentiles. The uncertainty show similar features to the second frequentist approach. The correlation matrix (panel c) is showing milder correlations with respect to the normalization parameters. The data without reported systematic uncertainty are normalized to the best fit values. The data legend is reported in Fig.~\ref{fig:lit}.}
\end{figure*}

To solve all the aforementioned issues and find an alternative way of estimating the uncertainty, while still using a frequentist minimizer, we developed a different procedure. Since the previous method relied on the computation of the covariance matrix to estimate the uncertainties, which could be inaccurate or approximate in cases where the parameter space is complicated \cite{minuit}, we attempted to obtain an error estimate by circumnavigating its calculation and using a hybrid approach. First, the normalization parameters, $n_{j}$, were sampled from a log-normal distribution, with the width given by systematic uncertainty of each dataset, which are equivalent of Bayesian priors. However, differently to what is done in the pure Bayesian case, as will be described in the following, the normalizations of each dataset were fixed to the sampled value, apart from those of \cite{rolfs1974,vogl1963,baily1950,lamb1957} ones, where no systematic uncertainty was reported and, finally, the minimization was performed. In this way, the normalizations are not anymore treated as fixed but unknown quantities, as occurs in purely frequentist approaches, but as random variables. The choice of log-normal distribution with respect to a classic gaussian relied on the fact that it describes the product of several random variables, thus, is more adequate for describing the systematic uncertainties coming from different sources. This procedure was repeated 1000 times to obtain a statistically relevant number of samples. The cross section distribution was then calculated by calculating an average of all the results from each sample, thus implicitly assuming a uniform weight for each sample. The resulting residuals can be seen in Fig.~\ref{fig:norm-1}~-~\ref{fig:norm-3}, alongside the correlation matrix. This method seems to give a more realistic uncertainty: where the discrepancy between the data is higher and less data are present, the extrapolation distribution is wider, whereas where more precise data exists the extrapolation uncertainty is smaller. The information about the systematics of each dataset is naturally included by creating virtual datasets with different normalizations according to their distributions. Additionally, the correlation matrix is not governed anymore by the normalization parameters. Instead, it shows a more clear connection between the different R-matrix parameters. This way of estimating the uncertainty comes straightforwardly from the sampling of the data normalizations, thus can be easily computed even if the parameter space is complicated and if the Hessian estimation can not be applied. Additionally, it can be easily adapted to add covariant systematic contributions for the absolute scale of the datasets, like the stopping power one which is common across the different datasets, and the uncertainty on the energy calibration of each experiment, and could be pursued in the future evaluations. The major drawback is that it relies on the reliability of the reported systematic error estimation of each study.


\subsection{Bayesian Approach}

This approach is build upon the Bayes' theorem, which describes the probability of a parameter based on prior knowledge of its value. More precisely:

\begin{align}
    P(A|B) &= \frac{P(B|A) P(A)}{P(B)},
\end{align}

where $P(A|B)$ is the posterior probability, $P(B|A)$ the likelihood, $P(A)$ the prior probability and $P(B)$ the marginal likelihood. The theorem permits to update beliefs about the probability of a quantity $A$, which at the beginning is enclosed in its prior $P(A)$, based on new evidence $B$. In case of the R-matrix, we want to update the distribution of all the parameters based on the experimental evidence of the cross sections, which can be done through the use of the likelihood function. Thus, in case of the Bayesian method, the parameters are treated as random variables, whereas in the case of frequentist methods these are fixed and unknown quantities. In the former case, the attention is put on probability distribution that reflects our uncertainty about their true value before and after seeing data, that is obtained by maximizing the posterior distribution. In the latter case, instead, the goal is to estimate these fixed parameters using observed data and maximizing the likelihood function.

To construct the Bayesian framework, we started by defining the prior distributions that encapsulate our beliefs about the parameters before observing the data. The distributions of radiative and proton widths, alongside the energies of the resonances, were set as uniform distributions i.e.\ uninformative priors, since no value is preferred over the other. For the normalization parameters of each dataset, log-normal distributions were selected with the width given by the reported systematic uncertainty. When no systematic uncertainty was reported in a study, an uninformative prior was used instead. The prior of the ANC for the capture to the ground state was considered Gaussian since its value was independently measured~\cite{artemov2022}. Then, the logarithm of the likelihood function, $\mathcal{L}$, was defined as~\cite{dagostini1994,bohm2017}:

{\footnotesize
\begin{align}
    \log{\mathcal{L}} &= \frac{1}{2}\sum_{i,j} \left[ -\log{\left(2\pi\sigma_{\textup{stat,i}}^{2}\right)} - \left(\frac{n_{j}y_{\textup{obs,i}} - y_{\textup{theo},i}}{n_{j}\sigma_{\textup{stat},i}} \right)^{2} \right].
\end{align}}

\begin{figure}
    \centering
    \includegraphics[width=0.95\linewidth]{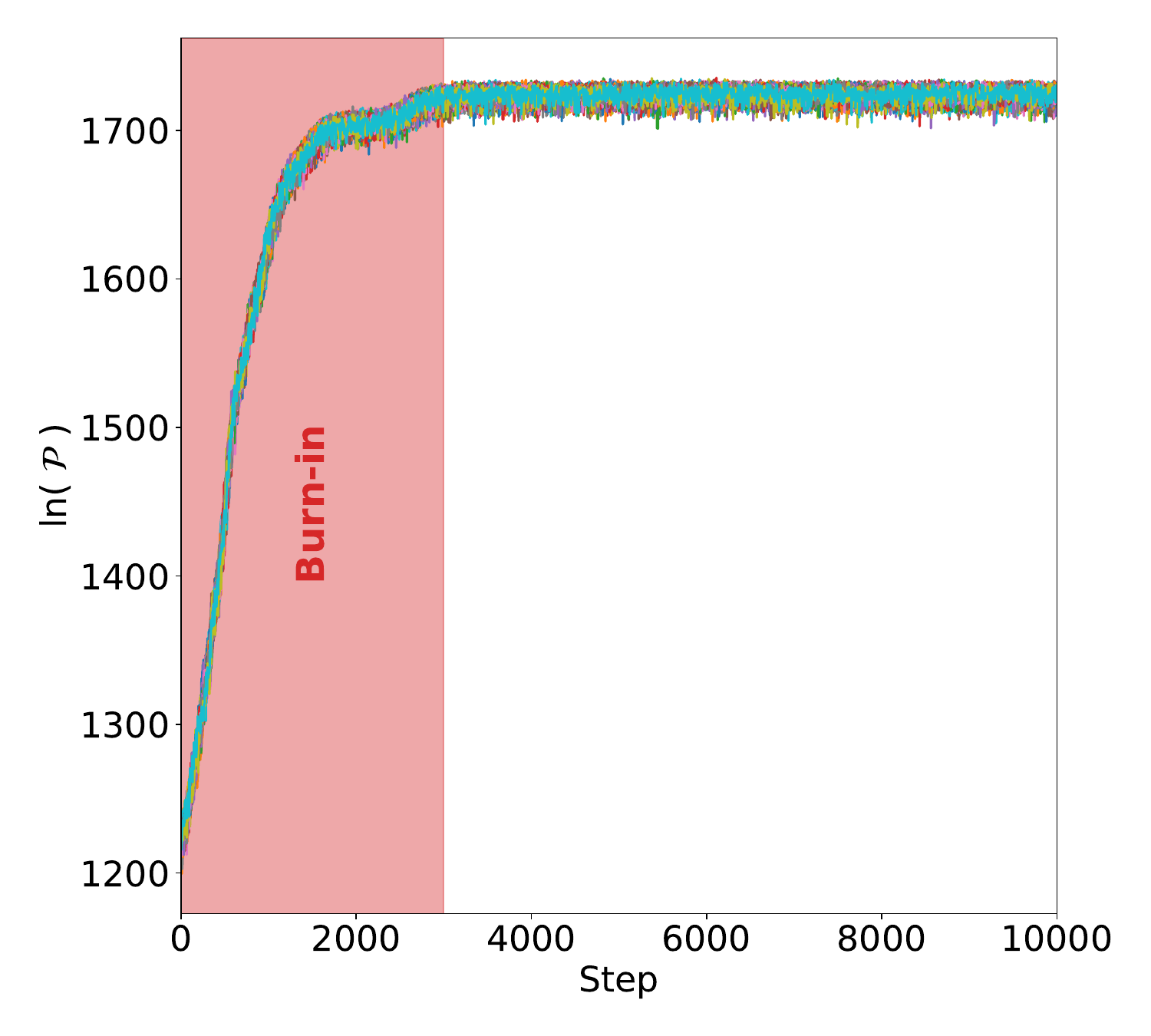}
    \caption{The log posterior probabilities for each MCMC chain. The initial burn-in region is highlighted, after which each chain reaches the minimum and draws samples from it.}
    \label{fig:burn}
\end{figure}

Since directly computing the posterior distribution is unfeasible, due to the complexity of the integral involved in the marginal likelihood $P(B)$, we used the Markov Chain Monte Carlo (MCMC) method \cite{press2002}. This is a class of algorithms used to sample the posterior distribution of model parameters. These algorithms generate a sequence of samples from the desired distribution, called chains, by constructing a Markov Chain that has the posterior distribution as its equilibrium one. Thus, it is possible to generate a sequence of samples that, after a "burn-in" period i.e.\ the number of samples after which the Markov Chain reaches the minimum, are drawn from the posterior distribution. The Metropolis-Hastings algorithm is usually used to achieve that. The algorithm proposes new parameters for the chain based on a proposal distribution and accepts or rejects these states based on an acceptance criterion to ensure convergence to the target distribution. The choice of the proposal distribution is crucial as it directly affects the efficiency and convergence of the Markov Chain. In case of the R-matrix fitting, we selected the approach followed by the \textit{emcee} package~\cite{emcee}, which is one of the most tested and used tool for the MCMC analysis. It takes care of instantiating several parallel MCMC chains, which explore the parameter space, and draws, after the initial burn-in period, the samples from the posterior distribution. The number of parallel chains was set to twice the number of free parameters to avoid auto-correlation issues \cite{emcee}. Each sampler was run from a random position drawn from the priors. In total, 10,000 samples were drawn for each chain. The observed burn-in period was 2,000 samples, as can be seen in Fig.~\ref{fig:burn}. The uncertainty and the parameter distribution were then calculated from the last 1000 samples, where the minimum was clearly reached. The resulting residuals and the covariance matrix are shown in Fig.~\ref{fig:emcee-1}~-~\ref{fig:emcee-3}. Noted that the uncertainty obtained with this approach shows no structure and is rather constant throughout the spanned energy region, very similar to the results of the first frequentist approach. Additionally, the covariance matrix shows values close to unity for the normalization parameters. This points out the fact that the classic MCMC method seems to struggle in case the parameter space shows signs of multi-modality, as in the case under study. The chains are easily stuck in one minimum and are not able to sample correctly the entirety of the parameter space. Thus, the total uncertainty of the extrapolation can be underestimated in cases where multiple minima are present in the parameter space since the posterior distribution can not be adequately sampled, even if the convergence is reached as pointed by Fig.~\ref{fig:burn}.

\begin{figure}
    \centering
    \includegraphics[width=\linewidth]{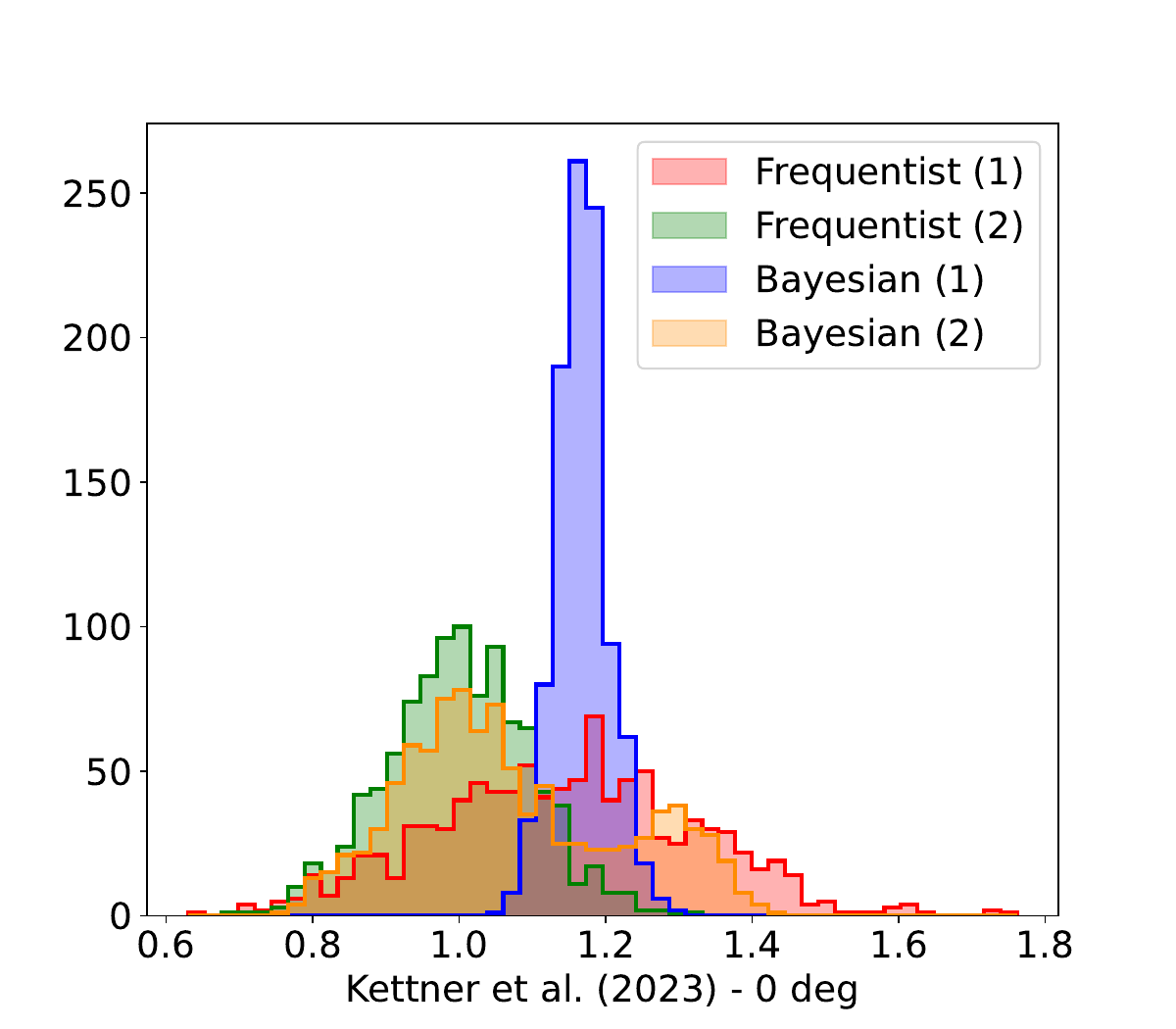}
    \caption{The distributions of the \citet{kettner2023} dataset at 0 deg for the different uncertainty estimations. In case of the Bayesian (2), two different modes are found which are not evident in the Bayesian (1) approach nor in the frequentist approach.}
    \label{fig:kettner}
\end{figure}

Due to the similarity of the results to those from the first frequentist method, we decided to explore more in depth its possibilities with a more advanced MCMC procedure, called the parallel tempering MCMC method~\cite{vousden2016}. This is an algorithm designed for running multiple chains with different acceptance distributions. The chains with flatter functions allow a system to overcome probability barriers more easily, e.g.\ when the system is stuck in a local minimum. The acceptance functions are frequently switched between the parallel chains during the sampling. In this way, the flatter chains are better at finding local minima and exploring the whole parameter space, whereas the other chains are better at exploring each minimum. This is the usual procedure to explore multi-modal parameter spaces in which the classic MCMC method gets stuck in one minimum instead of exploring the entirety of the parameter space. To perform the sampling in context of the R-matrix, 20 different temperatures were selected following the prescriptions from~\cite{vousden2016}. For each temperature, the procedure was equivalent to that previously described. The final residuals, alongside the correlation matrix, are shown in Fig.~\ref{fig:ptemcee-1}~-~\ref{fig:ptemcee-3}. This approach shows results very close to those of the second frequentist approach. Both the uncertainty shows more structure and the correlation between the normalization parameters are much more relaxed. Additionally, this MCMC algorithm is able to overcome the probability barriers between the modes of the different minima and explore the parameter space much more efficiently. An example of this can be seen in Fig. \ref{fig:kettner} showing the normalization parameter for the \citet{kettner2023} study. In case of the tempered samples, much wider posterior is sampled, which increase the uncertainty in the region of these data. In the other Bayesian case, one pronounced minimum is found, where the sampler is stuck. This is not a problem for the hybrid methodology method since the usage of the normalization factors is quite different in that case, as described before, neither for the first frequentist approach, since the uncertainties of the data are being inflated. Finally, it is important to note that alternatives to parallel tempering method exists, as Hamiltonian Monte Carlo~\cite{wang2019} and Importance Sampling~\cite{tokdar2010}, that could be investigated in the future.

\begin{figure}
    \centering
    \includegraphics[width=\linewidth]{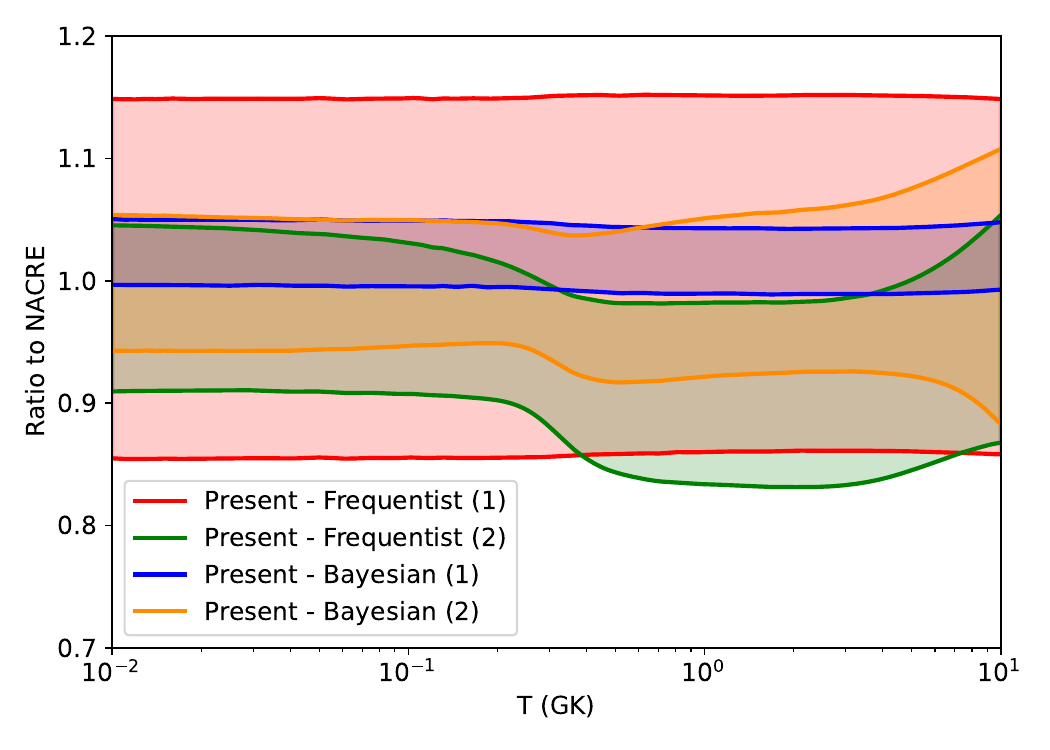}
    \caption{The comparison of the calculated rates with the four different approaches relative to the rate from the NACRE \cite{angulo1999} compilation.}
    \label{fig:rate}
    \vspace{0.5cm}
    \includegraphics[width=\linewidth]{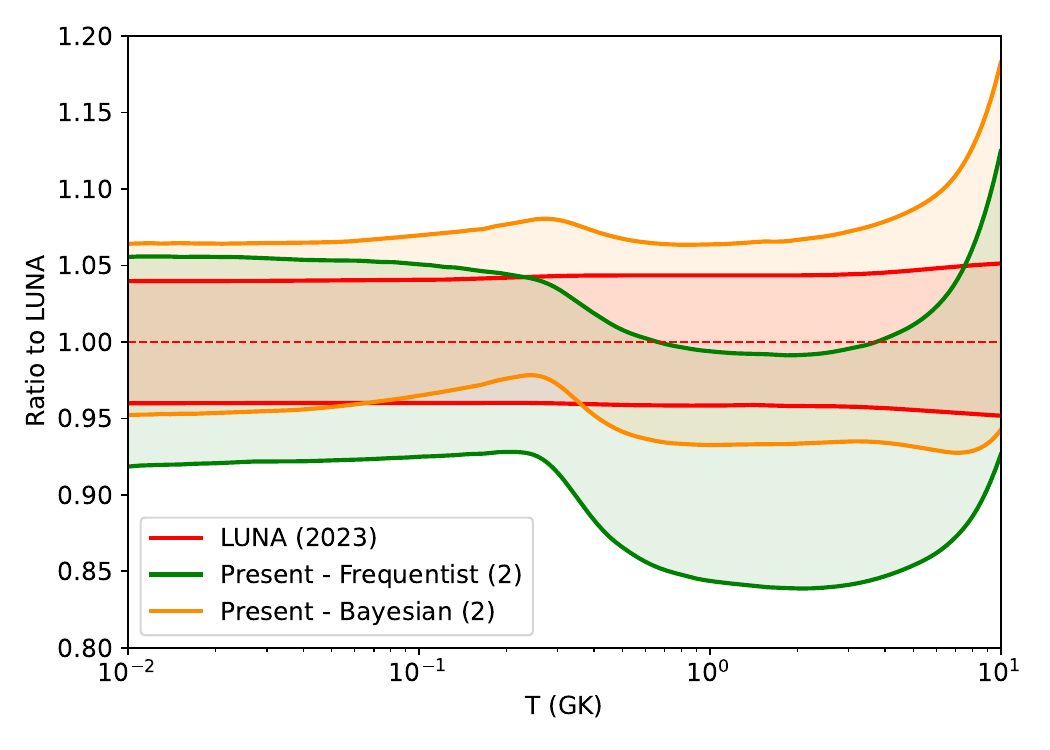}
    \caption{The comparison of the calculated rate with respect to the rate from LUNA \cite{skowronski2023_prl}. The frequentist and Bayesian uncertainties refers to the second procedure of each illustrated approach.}
    \label{fig:rate-latest}
\end{figure}

Once the cross section distributions were obtained from all the procedures, the astrophysical rates were calculated, using:

\begin{equation}
    \left< \sigma v \right>=\left ( \frac{8}{\pi\mu} \right )^{1/2}\left ( \frac{1}{kT} \right )^{3/2}\int_{0}^{\infty}E\sigma(E)e^{-{E/kT}}\mathrm{d}E,
\end{equation}

where $\sigma$ and $\mu$ are the cross section and reduced mass of the projectile-target system, $T$ the stellar temperature, and $k$ the Boltzmann constant. The comparison of the rates between the different approaches is shown in Fig.~\ref{fig:rate}, while in Fig.~\ref{fig:rate-latest} the second frequentist and the second Bayesian methods are confronted with the LUNA rate \cite{skowronski2023_prl}. The former clearly shows the facts that the first frequentist and Bayesian approaches overestimate and underestimate the uncertainties, whereas the other two are more consistent with each other. The comparison with the LUNA rate shows that the new rates are in agreement with the latest results, even though the uncertainties are slightly different and the rate is higher towards the \SI{10}{GK}, due to the availability of the \citet{kettner2023} data, which were not used in the LUNA \cite{skowronski2023_prl} evaluation.

\section{\label{sec:conclusions}Conclusions}

We performed an R-matrix analysis of the ${}^{12}$C\pg{}${}^{13}$N reaction including all the latest measurements. Due to discrepancies in the data, it was the perfect case to test the robustness of different frequentist and Bayesian approaches. The fitting of the data against the R-matrix model and the uncertainty estimation was performed in four different ways: one frequentist, one hybrid and two Bayesian. The uncertainty estimation can be rather troublesome in cases where discrepant data are present. In particular, in the case of the frequentist approach the Hessian estimation can give rise to very small uncertainties, which must be then inflated to give reasonable results. Additionally, the addition of the normalization parameters can give rise to different comparable local minima. These parameters can create ambiguities if an improper minimization algorithm is used. An alternative approach, which relies on sampling the datasets normalizations and perform parallel minimizations, seems to give more reliable uncertainty and overcome the problem of the covariance matrix calculation. In case of the Bayesian procedure, the standard MCMC approach could lead to the underestimation of the uncertainties due to the fact that these algorithms tend to stuck in a single mode. The tempered MCMC approach, instead, is able to solve this issue by introducing the temperature of the chain, which makes it more likely to overcome the single-modes and explore more efficiently the entire parameter space. By applying these improvements, both the frequentist and the Bayesian approaches give consistent results and can be applied to variety of cases. New reaction rates were calculated which were compatible to those published in the latest LUNA reaction studies.

\begin{acknowledgments}
D. Ciccotti and the technical staff of the LNGS are gratefully acknowledged for their indispensable help.
Financial support by INFN,
the Italian Ministry of Education, University and Research (MIUR) through the ``Dipartimenti di eccellenza'' project ``Physics of the Universe'',
the European Union (ERC-CoG \emph{STARKEY}, no. 615604; ERC-StG \emph{SHADES}, no. 852016; and \emph{ChETEC-INFRA}, no. 101008324),
Deut\-sche For\-schungs\-ge\-mein\-schaft (DFG, BE~4100-4/1),
the Helm\-holtz Association (ERC-RA-0016),
the Hungarian National Research, Development and Innovation Office (NKFIH K134197, FK134845),
the European Collaboration for Science and Technology (COST Action ChETEC, CA16117)
is gratefully acknowledged.
M.A., C.G.B, T.D., and R.S.S. acknowledge funding from STFC (grant ST/P004008/1).
J.S., G.I. and A.Bo. thank R.\,J.\ deBoer for his support with AZURE2.
We thank the referee for his/her comments during the review process. 
For the purpose of open access,  authors have applied a Creative Commons Attribution (CC BY) license to any Author Accepted Manuscript version arising from this submission.
\end{acknowledgments}

\appendix


\providecommand{\noopsort}[1]{}\providecommand{\singleletter}[1]{#1}%

\end{document}